\documentclass[]{aastex63}

\usepackage{tabularx}
\usepackage{color}
\usepackage{amsmath}

\shorttitle{Resolving Io's Volcanoes with the LBT}
\shortauthors{de Kleer et al.}

\graphicspath{{./}{figures/}}

\begin{document}

\title{Resolving Io's Volcanoes from a Mutual Event Observation at the Large Binocular Telescope}

\correspondingauthor{Katherine de Kleer}
\email{dekleer@caltech.edu}

\author[0000-0002-9068-3428]{Katherine de Kleer}
\affiliation{California Institute of Technology \\
1200 E California Blvd M/C 150-21 \\
Pasadena, CA 91125}

\author{Michael Skrutskie}
\affiliation{Department of Astronomy, University of Virginia, Charlottesville, VA}

\author{Jarron Leisenring}
\affiliation{Steward Observatory, University of Arizona, Tucson, AZ}

\author{Ashley G. Davies}
\affiliation{Jet Propulsion Laboratory -- California Institute of Technology, Pasadena, CA}

\author{Al Conrad}
\affiliation{The Large Binocular Telescope Observatory, University of Arizona, Tucson, AZ}

\author{Imke de Pater}
\affiliation{Department of Astronomy, University of California at Berkeley, Berkeley, CA}

\author{Aaron Resnick}
\affiliation{}

\author{Vanessa P. Bailey}
\affiliation{Jet Propulsion Laboratory -- California Institute of Technology, Pasadena, CA}

\author{Denis Defr\`ere}
\affiliation{Institute of Astronomy, KU Leuven, Leuven, Belgium}

\author{Phil Hinz}
\affiliation{The University of California at Santa Cruz, Santa Cruz, CA}

\author{Andrew Skemer}
\affiliation{The University of California at Santa Cruz, Santa Cruz, CA}

\author{Eckhart Spalding}
\affiliation{Department of Physics, University of Notre Dame, Notre Dame, IN}

\author{Amali Vaz}
\affiliation{Steward Observatory, University of Arizona, Tucson, AZ}

\author{Christian Veillet}
\affiliation{The Large Binocular Telescope Observatory, University of Arizona, Tucson, AZ}

\author{Charles E. Woodward}
\affiliation{Minnesota Institute for Astrophysics, University of Minnesota, Minneapolis, MN}

\begin{abstract}

Unraveling the geological processes ongoing at Io's numerous sites of active volcanism requires high spatial resolution to, for example, measure the areal coverage of lava flows or identify the presence of multiple emitting regions within a single volcanic center. In \cite{dekleer2017} we described observations with the Large Binocular Telescope (LBT) during an occultation of Io by Europa at $\sim$6:17 UT on 2015 March 08, and presented a map of the temperature distribution within Loki Patera derived from these data. Here we present emission maps of three other volcanic centers derived from the same observation: Pillan Patera, Kurdalagon Patera, and the vicinity of Ulgen Patera/PV59/N Lerna Regio. The emission is localized by the light curves and resolved into multiple distinct emitting regions in two of the cases. Both Pillan and Kurdalagon Paterae had undergone eruptions in the months prior to our observations, and the location and intensity of the emission is interpreted in the context of the temporal evolution of these eruptions observed from other facilities. The emission from Kurdalagon Patera is resolved into two distinct emitting regions separated by only a few degrees in latitude that were unresolved by Keck observations from the same month.

\end{abstract}

\section{Introduction} \label{sec:intro}

Io's multitude of rapidly-evolving volcanoes provide our best opportunity to directly study ongoing silicate volcanism on another world. High-resolution visible imagery from the \textit{Voyager} and \textit{Galileo} missions revealed that many of Io's volcanic centers are complex systems of paterae, fissures, lava flows, and volatile deposits, with recent activity often superposing flows and deposits over relics of past activity \citep[e.g.,][]{schaber1980,schaber1982,williams2011,williams2011map}. Detailed imagery of these regions can help piece together the history of activity and the range of volcanic styles active within a given region, which in turn provide clues into magma composition and vent conditions.

Specific properties of Io's volcanic activity, such as the magma temperature and the distribution of volcanoes across the surface, have the potential to inform our understanding of the tidal heating process that ultimately drives Io's activity. However, the spatial distribution of Io's volcanoes is not a good match to current tidal heat deposition models \citep{hamilton2013,davies2015,dkdp2016b}. This disconnect may result from inhomogeneities in Io's interior that lead to different amounts of melt generation in different areas; from lateral movement of melt within Io's interior that shifts the pattern of surface heat flow away from end-member model expectations \cite[e.g.][]{tyler2015,steinke2020}; or from processes taking place in the upper mantle and crust that erase the patterns of heat deposition at depth such that the volcanic heat flow at a given surface location does not mirror the tidal heat deposition directly below that location. Linking deep heat deposition patterns to their surface expressions is challenged by our nearly complete lack of knowledge about how magma is transported from the mantle to the surface, and about the properties of Io's magma reservoirs and plumbing systems.

High spatial resolution observations of Io's volcanic centers provide insight into the volcanic processes taking place at scales commensurate with the spatial resolution of the data. At spatially extended sources, this may result in a qualitative shift in our understanding of the volcanic processes at work \citep{dekleer2017}. For example, in data with resolution typical of spacecraft and adaptive optics (AO) on ground based telescopes ($\sim$100 to $>$500 km), the progression from localized fountaining to extensive lava flow emplacement during voluminous fissure eruptions must be inferred from the evolution of the unresolved emission spectrum \citep[e.g.,][]{davies1996,davies2001,depater2014outbursts}. Such eruptions have been known to rapidly emplace extensive lava flows covering hundreds of km$^2$ in a few days \citep{depater2014outbursts}. A higher spatial resolution of $\sim$10 km could directly distinguish the areas of fountaining from the emplacing lava flows, more tightly constraining the temporal and thermal evolution of the fountaining episode. This would allow for more robust modeling of both the fountaining and lava flow emplacement processes that could constrain the composition and properties of the lava.  Such detailed studies have only been possible to date using data from the closest of past spacecraft fly-bys \citep{williams2001}. 

Adaptive optics (AO) observations from 8-10 m telescopes match, and in some cases even exceed, the resolution of the majority of past spacecraft observations of Io's near-infrared emission. AO data reveal emission from distinct volcanic centers on Io's disk and enable precise measurements of near-infrared emission from a few up to over a dozen active volcanoes on Io's surface in a single snapshot observation \citep[e.g.,][]{depater2004,marchis2005,conrad2015}. However, even this resolution ($\sim$40--100 mas, or $\sim$140--360 km at Io's geocentric distance) is only capable of distinguishing volcanoes from one another and not resolving different emitting centers within a given surface region; the latter requires either new instrumentation or new techniques. \cite{conrad2015} used the Large Binocular Telescope Interferometer (LBTI) to directly resolve emission from within Loki Patera, Io's most powerful persistent volcano, and found that the thermal emission arose from two emitting regions separated by 100-200 km in the south of the patera. In 2015, we used the LBTI to observe Io during an occultation by Europa, which allowed us to resolve thermal emission from Loki Patera at a spatial sampling of $\sim$2 km \citep{dekleer2017}. The temperature map derived from these data pointed to the presence of two resurfacing waves that originated in different parts of the patera and converged in the southeast corner a few months before the observations. This was the first high spatial resolution temperature map of the entire volcanically active area of Loki Patera (an area of over 21,500 km$^2$; \cite{veeder2011}). 

Loki Patera is an ideal target for such an analysis because it is among the largest and most powerful active sites on Io, and emission is confined to a distinct and well-characterized patera. However, the mutual occultation technique can also be used to resolve smaller and fainter thermal centers, where it yields the precise size and location of the emitting region and can resolve the emission into distinct emitting areas that would remain unresolved even using AO on a 10-m telescope.

Mutual satellite occultation events require the orbital plane of the satellites to be edge-on from the view of Earth, which only occurs for a few months every six years. However, even a single observation of such an event can provide high resolution emission maps of several of Io's volcanoes. Here we apply the techniques presented in \cite{dekleer2017} to an analysis of the three other hot spots that were observed during the same event with sufficient signal to extract a light curve. Section \ref{sec:obs} reviews the observations and data reduction techniques, including the specific details of these hot spots. The results are presented and discussed in Section \ref{sec:results} and conclusions are summarized in Section \ref{sec:conc}.

\section{Observations and Methods} \label{sec:obs}

\subsection{Occultation observations}

\begin{figure}[ht!]
\centering
\includegraphics[width=12cm]{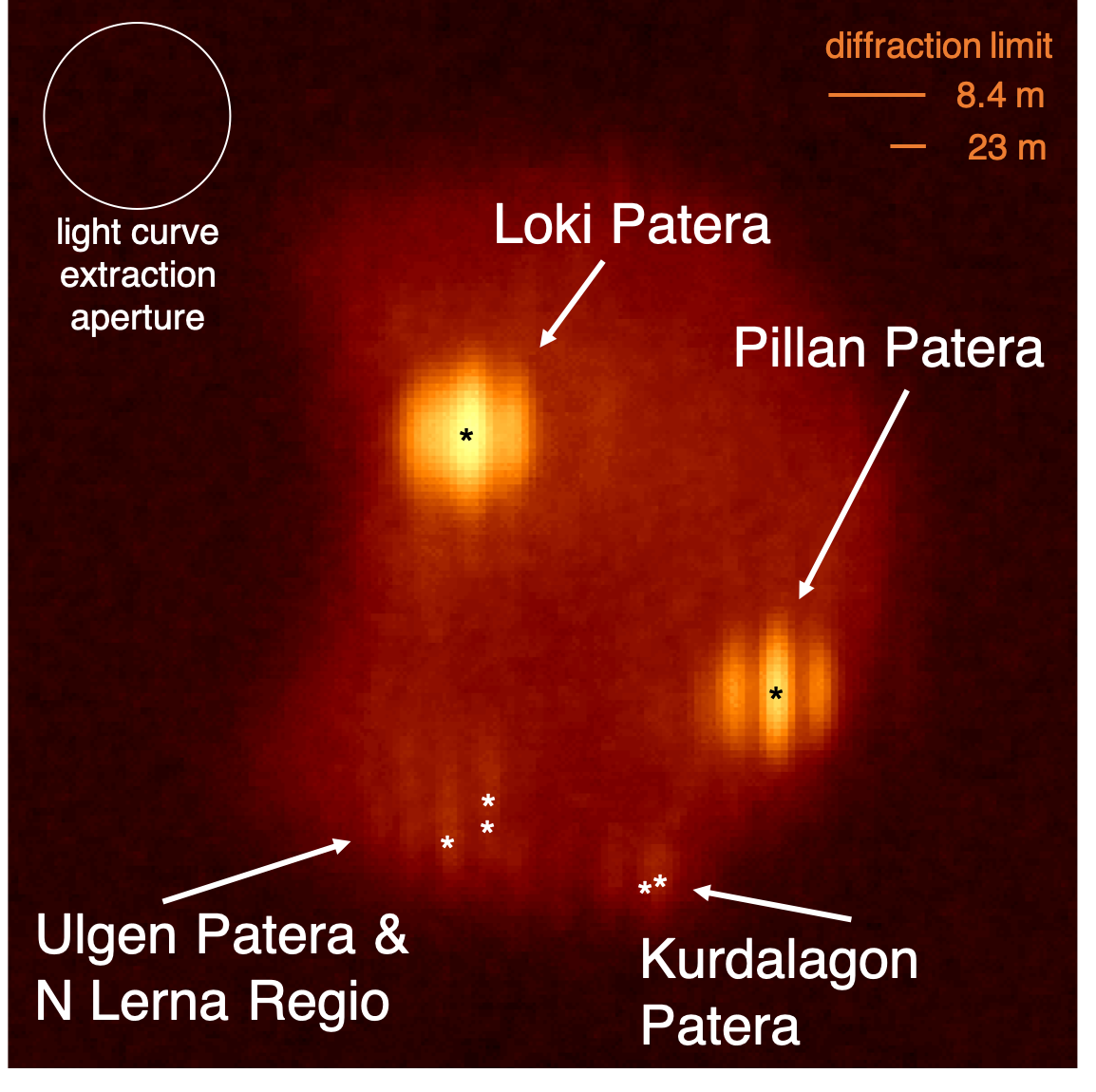}
\caption{Image of Io from LBTI during the occultation on 2015 March 08; Europa's limb can be seen covering a portion of Io's surface. The four hot spots discussed here are labeled, and the best-fit locations of emitting areas, as derived from the light curves, are shown with stars. The light curve extraction aperture has an angular radius of 140 mas, or $\sim$450 km at disk center. Diffraction limit scale bars are provided for context and are calculated at 4.8 $\mu$m for aperture sizes of 8.4 m (the diameter of each LBT mirror) and 23 m (the full baseline of the LBTI). \label{fig:obsim}}
\end{figure}

\begin{figure}[ht!]
\centering
\includegraphics[width=16cm]{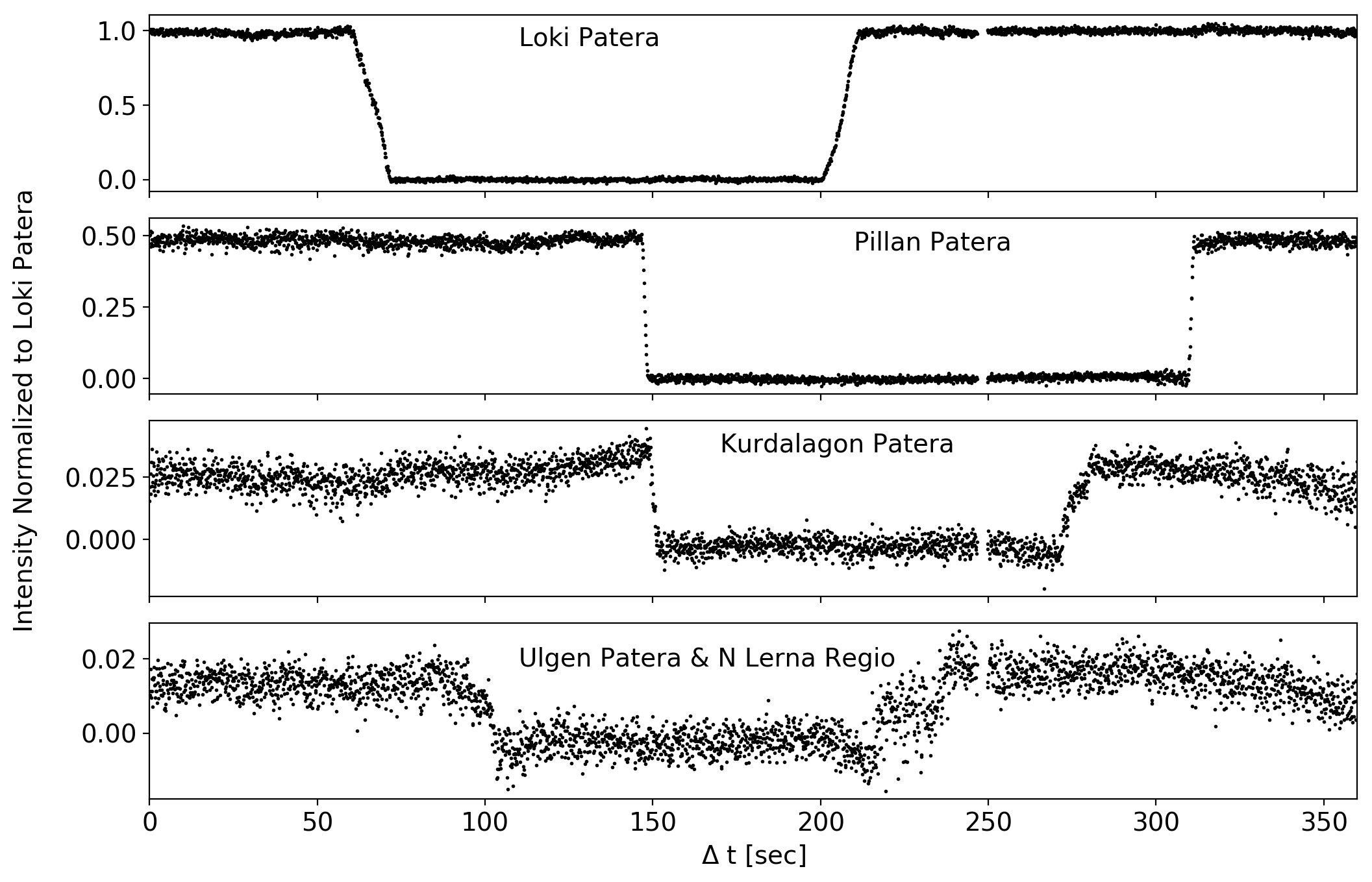}
\caption{Light curves of the four hot spots extracted from the LBTI mutual occultation dataset obtained on 2015 March 08. \label{fig:alllc}}
\end{figure}

We observed Io during an occultation by Europa on 2015 March 08 UT using Fizeau interferometric imaging with the LBTI \citep{hill2006,hinz2012} Mid-InfraRed camera (LMIRcam) \citep{leisenring2012,leisenring2014} with adaptive optics. Observations were acquired in M-band (4.6-5.0 $\mu$m) with an integration time of 15.4 ms and an average cadence of 120 ms. A sample view of Io and Europa during the occultation is shown in Figure \ref{fig:obsim}, and the four volcanoes discussed here are labeled. The observations and data reduction methods are identical to those described in \cite{dekleer2017} and are only briefly summarized here. The LMIRcam images were reduced using standard data reduction methods, including the subtraction of sky background estimated from background frames acquired close in time to the science frames. A model of reflected sunlight off Io's disk was subtracted frame-by-frame, yielding images containing only hot spot emission. Within each frame, the flux density of each hot spot was extracted via aperture photometry using an effective circular aperture with a radius of 140 mas. The light curves of four thermal hot spots were extracted via this method: Loki Patera (310$^{\circ}$W, 12$^{\circ}$N), Pillan Patera (244$^{\circ}$W, 12$^{\circ}$S), Kurdalagon Patera (219$^{\circ}$W, 50$^{\circ}$S), and combined emission from 2-3 regions in the vicinity of Ulgen Patera (288$^{\circ}$W, 41$^{\circ}$S) and N Lerna Regio (around 290$^{\circ}$W, 55$^{\circ}$S). The extracted light curves of all four of these volcanic regions are shown in Figure \ref{fig:alllc}, and the exact ingress and egress times are given in Table \ref{tbl:times}. The duration of ingress and egress ranged from $\Delta t$=1.8 to 24 seconds, during each of which $N_{frames}=$16 to 201 individual frames were obtained and used in the analysis. Individual datapoints have a signal to noise ratio (SNR) of 23, 6, and 4 for Pillan Patera, Kurdalagon Patera, and Ulgen Patera/N Lerna Regio respectively. Averaging over 5 seconds of clock time, the SNR increases to 144, 35, and 27 respectively. Analysis of the light curve of Loki Patera was the subject of \cite{dekleer2017}; here we present an analysis of the other three light curves. 

Flux calibration was performed by calibrating to the standard star HD 81192, using a Phoenix model template for a G7III type star \citep{stsci2013} normalized to the 2MASS and WISE measurements. The stellar flux density was extracted using the same aperture applied to Io; the fraction of the flux density captured by this aperture varied by 2\% between stellar observations, and we adopt this as the uncertainty arising from the choice of aperture. \cite{dekleer2017} found a flux density of 950$\pm$25 mJy for Loki Patera by calibrating directly to the star. The uncertainties include uncertainties in the flux ratio between Loki Patera and the star; uncertainties based on the 2MASS and WISE errorbars; and possible mismatches on the stellar template. The flux densities of the other volcanoes are measured relative to Loki Patera (see Figure \ref{fig:alllc}), and the uncertainty propagated. The flux density uncertainties are lower than derived for typical AO observations because of the very stable conditions on the night of observation, combined with improvement enabled by the modeling and removal of scattered sunlight from Io's disk prior to hot spot photometric extraction. 

The flux density of Pillan Patera is measured at
380$\pm$15 mJy, Kurdalagon Patera at 24$\pm$1 mJy, and Ulgen Patera$+$N Lerna Regio at 13$\pm$0.5 mJy. Converting to units of GW/$\mu$m/sr and applying a geometric correction for projected area by dividing flux density of each hot spot by the cosine of the emission angle, we measure flux densities of 36$\pm$2, 2.2$\pm$0.1 and 3.7$\pm$1.0 GW/$\mu$m/sr for Pillan Patera, Kurdalagon Patera, and Ulgen Patera$+$N Lerna Regio respectively. The flux densities of Pillan Patera and Kurdalagon Patera will be shown in the context of measurements from other telescopes during the same time period in Section \ref{sec:results}.

\begin{table}
\caption{Occultation Ingress/Egress Times on 2015 March 08 UT \label{tbl:times}}
\centering
\begin{tabular}{lcccccc}
\hline \hline
 & $t_{start}$ [UT] & $t_{end}$ [UT] & $\Delta t$ [sec] & $N_{frames}$ & Limb Advance Rate [km/s]$^a$ \\
\hline
\multicolumn{6}{l}{\textbf{Pillan Patera}} \\
\textit{Ingress} & 06:17:49 & 06:17:51 & 1.8 & 16 & 32 \\
\textit{Egress} & 06:20:32 & 06:20:34 & 2.1 & 18 & 31 \\
\hline
\multicolumn{6}{l}{\textbf{Kurdalagon Patera}} \\
\textit{Ingress} & 06:17:52 & 06:17:56 & 3.9 & 33 & 71 \\
\textit{Egress} & 06:19:53 & 06:20:03 & 9.8 & 81 & 15 \\
\hline
\multicolumn{6}{l}{\textbf{Ulgen Patera/PV59/N Lerna Regio}} \\
\textit{Ingress} & 06:16:47 & 06:17:07 & 19 & 156 & 21 \\
\textit{Egress} & 06:18:56 & 06:19:21 & 24 & 201 & 17 \\
\hline
\end{tabular}\\
\footnotesize{$^a$Speed of Europa's limb advancing across each location, in km on Io's surface per second}
\end{table}

\subsection{Light curve modeling}

\begin{figure}[ht!]
\centering
\includegraphics[width=16cm]{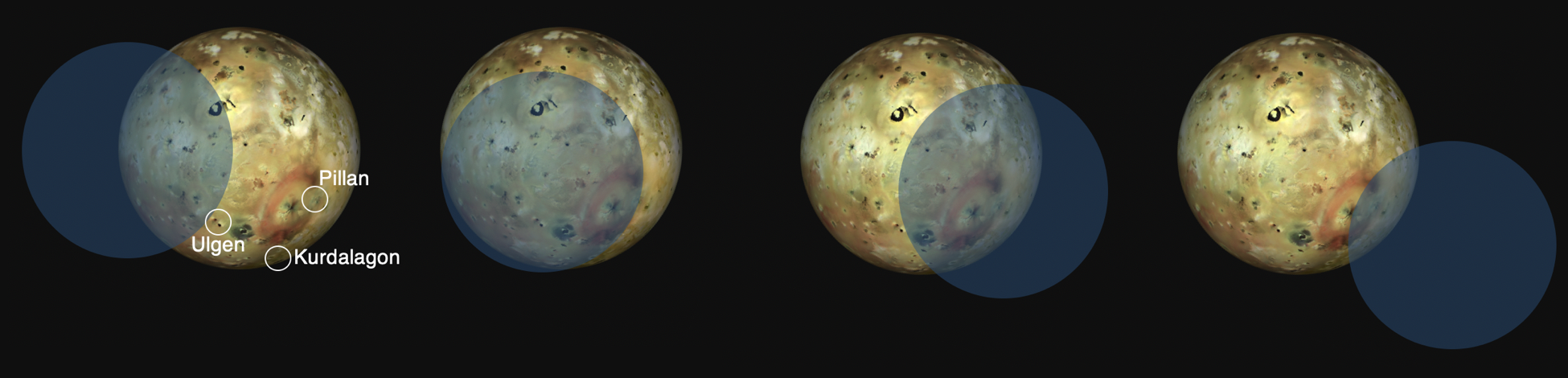}
\caption{Schematic showing the approximate position of Europa relative to Io at selected points during the occultation close to the ingress and egress times of the three hot spots analyzed in this paper. \label{fig:occseq}}
\end{figure}

The covering and uncovering of slivers of Io's surface by Europa's limb during ingress and egress leads to a decrease and then increase in the integrated flux density of each hot spot. The change in hot spot brightness during each timestep can thus be used to reconstruct the spatial distribution of the emission along the direction of motion of Europa's limb. If ingress and egress follow sufficiently different directions of motion at the location of a given volcano, a two-dimensional brightness distribution can be reconstructed. In the observations presented here, all hot spots with the exception of Pillan Patera had sufficiently different ingress and egress directions to determine the brightness distribution along two distinct axes (shown schematically in Figure \ref{fig:occseq}). At Pillan Patera the ingress and egress followed similar directions, resulting in a stronger longitudinal than latitudinal constraint.

For each hot spot, the region that is the intersection between the swaths of Io's surface covered and uncovered during the ingress and egress is identified. A model brightness map is created by dividing this region up into model pixels, and a model light curve is computed from the brightness map following the methods of \cite{dekleer2017} and using the JUP365 ephemeris\footnote{ssd.jpl.nasa.gov}. Emission from outside the identified region is fixed at zero, and a minimization is performed on the chi-squared value of the model light curve. Two different modeling approaches are tested. In the ``unconstrained model'' approach, the value of each pixel within the emitting region is permitted to vary independently, while in the ``Gaussian model'' approach the brightness distribution is forced to be Gaussian. Between one and three Gaussian emitting areas are required to fit the light curves; when multiple Gaussians are used, the free parameters are the center position and width of each Gaussian and the relative flux densities of the different components. The outcome of this procedure is a best-fit light curve model and its corresponding surface flux density distribution, for each hot spot and modeling approach. 

For Pillan Patera, Europa's limb followed a similar direction during ingress and egress, and the unconstrained modeling approach either does not converge or converges to a local minimum very close to its initial parameter position. However, a single Gaussian emitting area provides an excellent fit to the light curve with only three free parameters (x,y coordinates and full width at half maximum, FWHM, of the emitting area), and we therefore present only a 1-Gaussian model for Pillan Patera. Models with additional emitting regions were tested, but the minimizations either pushed them off the edge of the model image or placed them on top of one other, indicating a preference for the emission from Pillan Patera to be strictly localized to an area a few tens of kilometers in size.

For Kurdalagon Patera and Ulgen Patera/N Lerna Regio, the unconstrained model has many degenerate solutions due to the lower SNR of the data, and the resulting model solution depends heavily on initial conditions. In addition, neither hot spot could be fit well with a single-Gaussian model, which is qualitatively apparent from the fact that each light curve exhibits at least two steps (see Section \ref{sec:results}). However, in both cases the light curve can be quite well fit with either two or three Gaussians; the Gaussian model has much fewer parameters, and these are well constrained by the modeling. For this reason the Gaussian models are the preferred models for all hot spots, but one example best-fit unconstrained model for Kurdalagon Patera will be shown alongside the Gaussian model results as a demonstration.

\subsection{Imaging observations}\label{sec:imaging}

Io was also observed with Fizeau imaging with the LBTI on 2015 February 05 UT, when there was no occultation occurring. The observations took place during an eruption at Kurdalagon Patera and about a month before the occultation dataset; we include the imaging observations here to provide context to the interpretation of the occultation data. Observations were acquired with LMIRcam and adaptive optics in both M-band (4.6-5.0 $\mu$m) and L-band (3.4-4.0 $\mu$m) over a 5-hour period. Only the images taken between 11-12 UT, when Kurdalagon was visible, are analyzed here. During that period, four observation epochs were obtained in each filter. Images are produced by combining 400 frames each with a 30 ms exposure time. All images are rotated so that Io north is up, based on the known instrument rotation and the sub-observer latitude/longitude and north polar angle retrieved from JPL's Horizons database\footnote{https://ssd.jpl.nasa.gov/horizons.cgi}.\par
\begin{figure}[ht!]
\centering
\includegraphics[width=16cm]{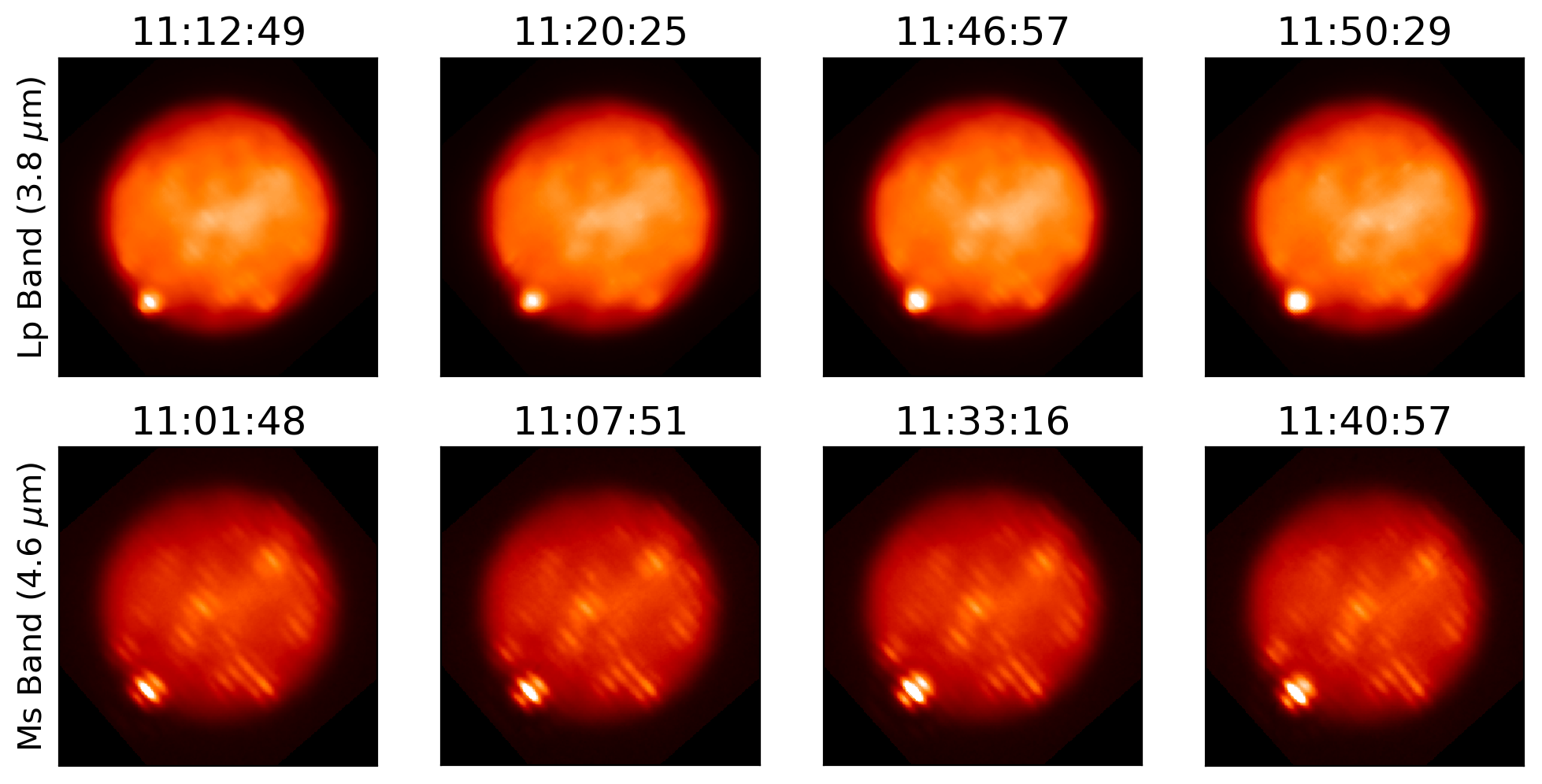}
\caption{Sequence of LBTI interferometric images of Io from 2015 February 05. Images are shown in two filters over a period of about an hour, and all panels within each row are on the same brightness scale. The bright volcano rotating onto Io's disk in the south is Kurdalagon Patera, which was undergoing an eruption at the time. The observed FWHM of the central fringe at Kurdalagon in the L-band images is broader than the FWHM of the PSF, indicating that the source is marginally resolved in the cross-fringe direction. Fewer volcanoes are seen at L than M band overall because Io's disk is brighter and because L-band is sensitive only to the higher-temperature emission. Io north is up in all images, and all times are given in UT on 2015 February 05.\label{fig:kurdims}}
\end{figure}

The images are flux calibrated assuming a fixed flux density of the background disk in regions without hot spots. The calibration flux density is derived from a set of calibrated observations from the Keck telescope in the same filters, adjusted for the difference in pixel size and orbital position at the time of each observation. This procedure is described in detail in \cite{dkdp2016a}. The flux density of Kurdalagon Patera in the LBTI images is extracted via aperture photometry. We use an aperture too small to include all the hot spot flux density in order to exclude nearby hot spots, and adjust the derived flux densities by a factor determined by applying the same aperture to observations of a point source and determining what fraction of the total flux density is missed. This technique was first suggested by \cite{gibbard2005}, and modified and used on Io's hot spots by \cite{depater2014} and \cite{dkdp2016a}. In this case, the point source is a star that is observed close in time to the target data for the purpose of point spread function (PSF) estimation: HD 81192 and HD 90718. All flux densities presented in this paper include a correction factor for geometric foreshortening calculated from the emission angle of the hot spot at the time of observation. This correction is appropriate for lava flows and lava lakes on flat surface regions with minimal topography; we note that significant topography or active lava fountaining would cause deviations from simple geometric foreshortening. The rotated and calibrated images are shown in Figure \ref{fig:kurdims}.
\section{Results \& Discussion} \label{sec:results}

\begin{table}
\caption{Best-fit Emission Components \label{tbl:results}}
\centering
\begin{tabular}{lcccc}
\hline \hline
 & Latitude & Longitude & Gaussian FWHM & Flux Density \\
 & [$^{\circ}$N] & [$^{\circ}$W] & [km] & [GW/$\mu$m/sr] \\
\hline
\textbf{Pillan Patera} & -12.0 & 243.7 & 34 & 36 \\
\hline
\multicolumn{5}{l}{\textbf{Kurdalagon Patera}} \\
Component 1 & -51.1 & 221.2 & 45 & 1.6 \\
Component 2 & -48.2 & 220.6 & 16 & 0.6 \\
\hline
\multicolumn{5}{l}{\textbf{Ulgen Patera/PV59/N Lerna Regio}} \\
Component 1 & -41.1 & 287.4 & 5 & 1.1 \\
Component 2 & -56.0 & 291.0 & 28 & 1.6 \\
Component 3 & -46.7 & 283.9 & 14 & 1.0 \\
\hline
\end{tabular}\\
\end{table}

\subsection{Pillan Patera}
\begin{figure}[ht!]
\centering
\includegraphics[width=10cm]{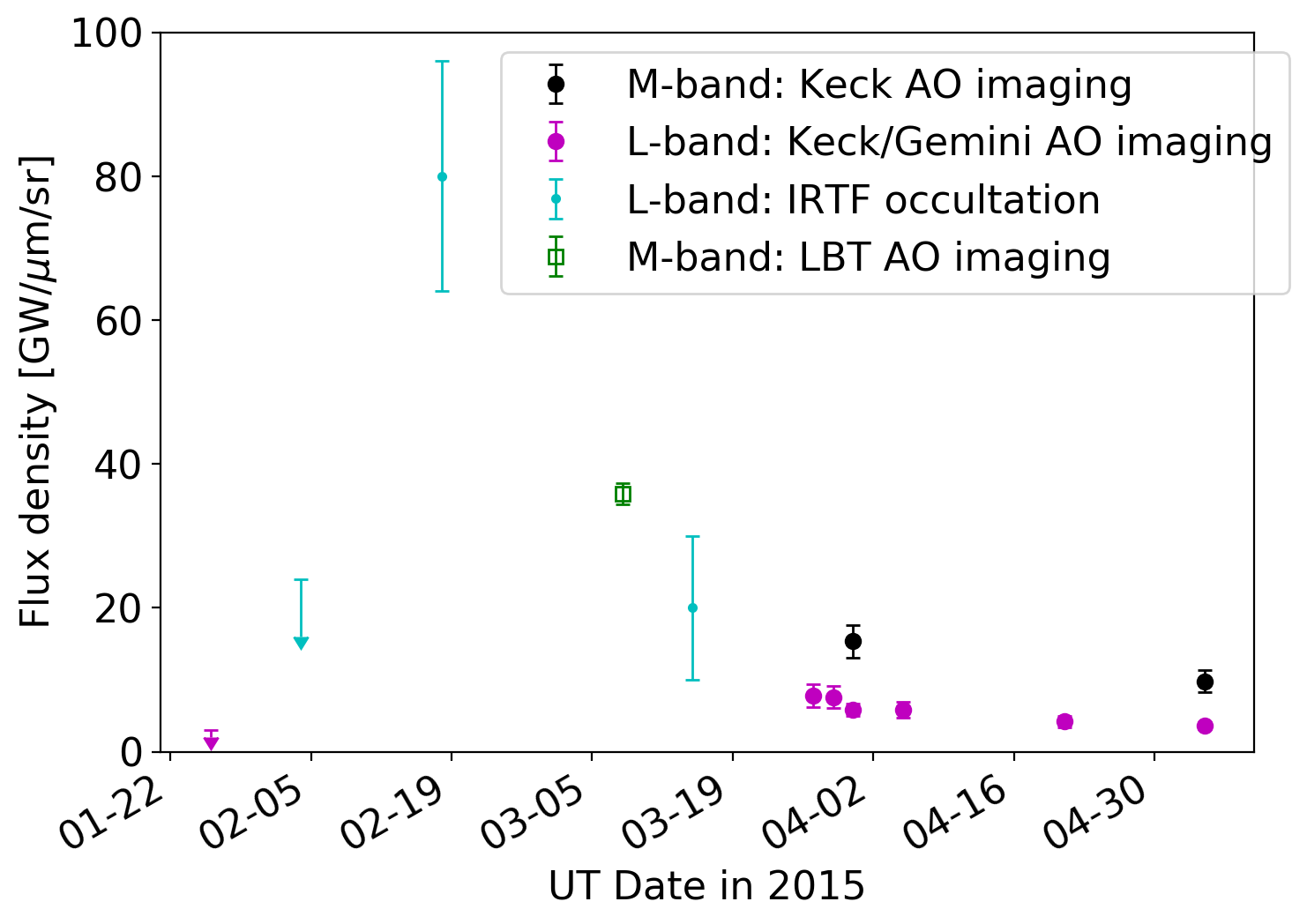}
\caption{Thermal emission from Pillan Patera over time in early 2015. The M-band LBT measurement is derived from the occultation data presented here, and all other measurements are taken from \cite{depater2016} and \cite{dkdp2016a}. \label{fig:pillan_timeline}}
\end{figure}

Multiple powerful volcanic eruption episodes have been observed at Pillan Patera since the first \textit{Galileo} observations in 1996 \citep{davies2001,depater2016}. This site underwent a major eruption a few months prior to the date of our observations. The eruption was first detected from the NASA IRTF 3.1-m telescope, from which a major spike in emission was seen on Feb 18, and could not have initiated before Feb 4 \citep{depater2016}. The LBT occultation data were taken 20 days after that first detection, and before the next detection of the eruption from any other telescope facility. The LBT measurement is placed in the context of the emission timeline from Jan-May 2015 in Figure \ref{fig:pillan_timeline}, and fits within the emission decay profile derived from data from other facilities. Note that M-band emission exceeds L-band emission by 30-50\% for typical temperatures of cooling Ionian lava flows ($\sim$300-600 K), but may be comparable to or even lower than the L-band emission for high temperatures such as are present during the eruption phase.

\begin{figure}[ht!]
\includegraphics[width=16cm]{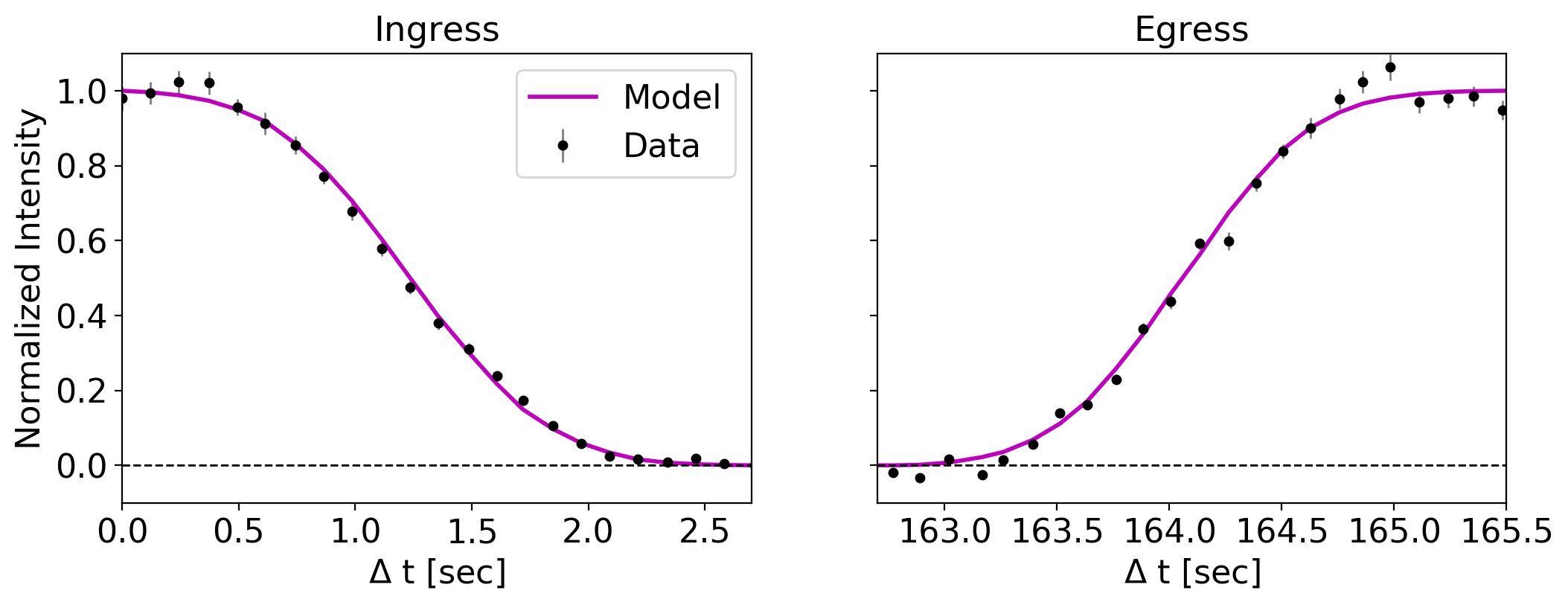}
\caption{Light curve of Pillan Patera with best-fit model, corresponding to the emission distribution shown in Figure \ref{fig:pillanim}. The model assumes one Gaussian emitting area; the free parameters in the fit are its location and width. \label{fig:pillanlc}}
\end{figure}

\begin{figure}[ht!]
\includegraphics[width=16cm]{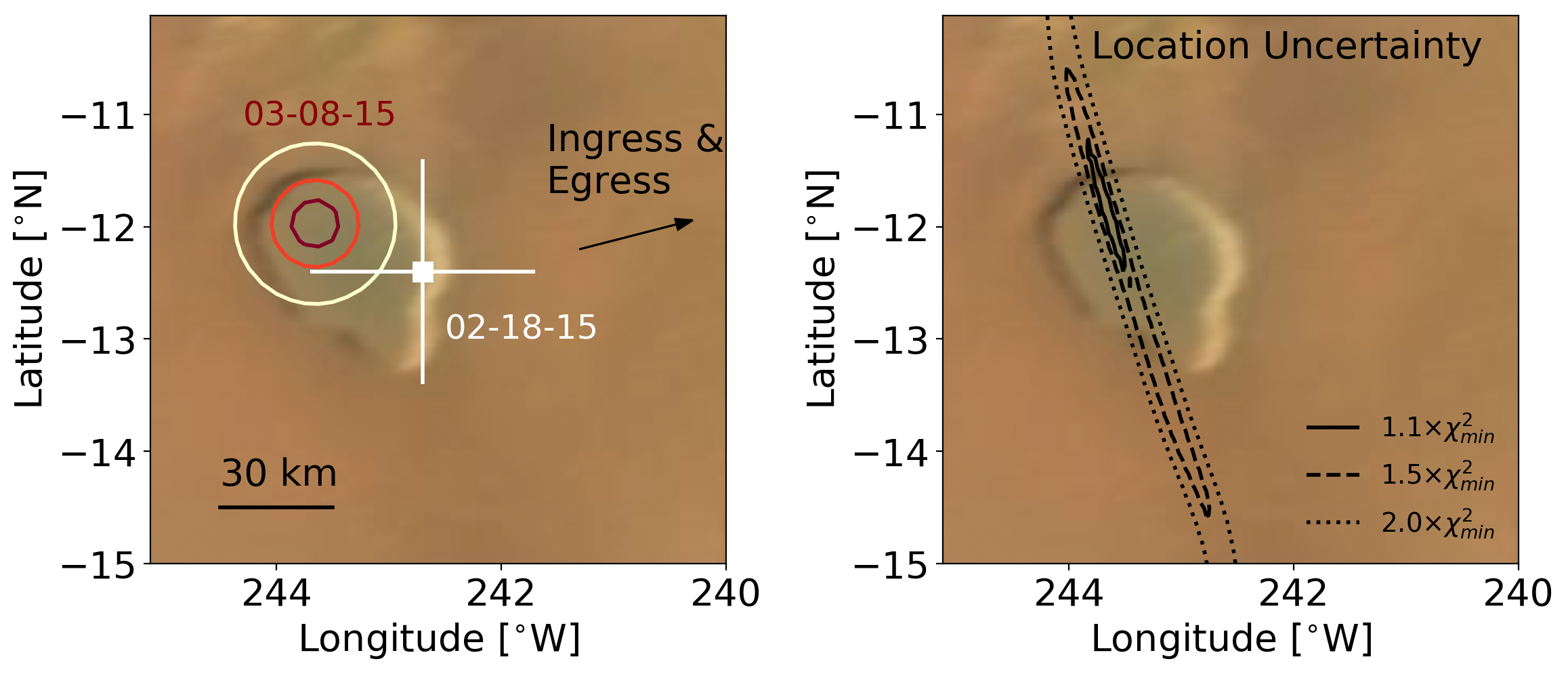}
\centering
\caption{[Left] Emission contours for Pillan Patera, corresponding to the model light curve shown in Figure \ref{fig:pillanlc}, overplotted on a USGS \textit{Voyager/Galileo} mosaic. The arrow indicates the approximate direction of movement of Europa's limb during ingress and egress, which at this location was nearly the same. The location of an eruption at Pillan Patera detected less than a month prior is shown for context \citep{depater2016}. Contours indicate 30, 70, and 90\% of peak emission. [Right] $\chi^2$ contours for peak emission assuming Gaussian emission with the Gaussian width fixed at the best-fit value. The location is constrained much more precisely along the direction of motion of Europa's limb during ingress and egress. \label{fig:pillanim}}
\end{figure}

\begin{figure}[ht!]
\includegraphics[width=10cm]{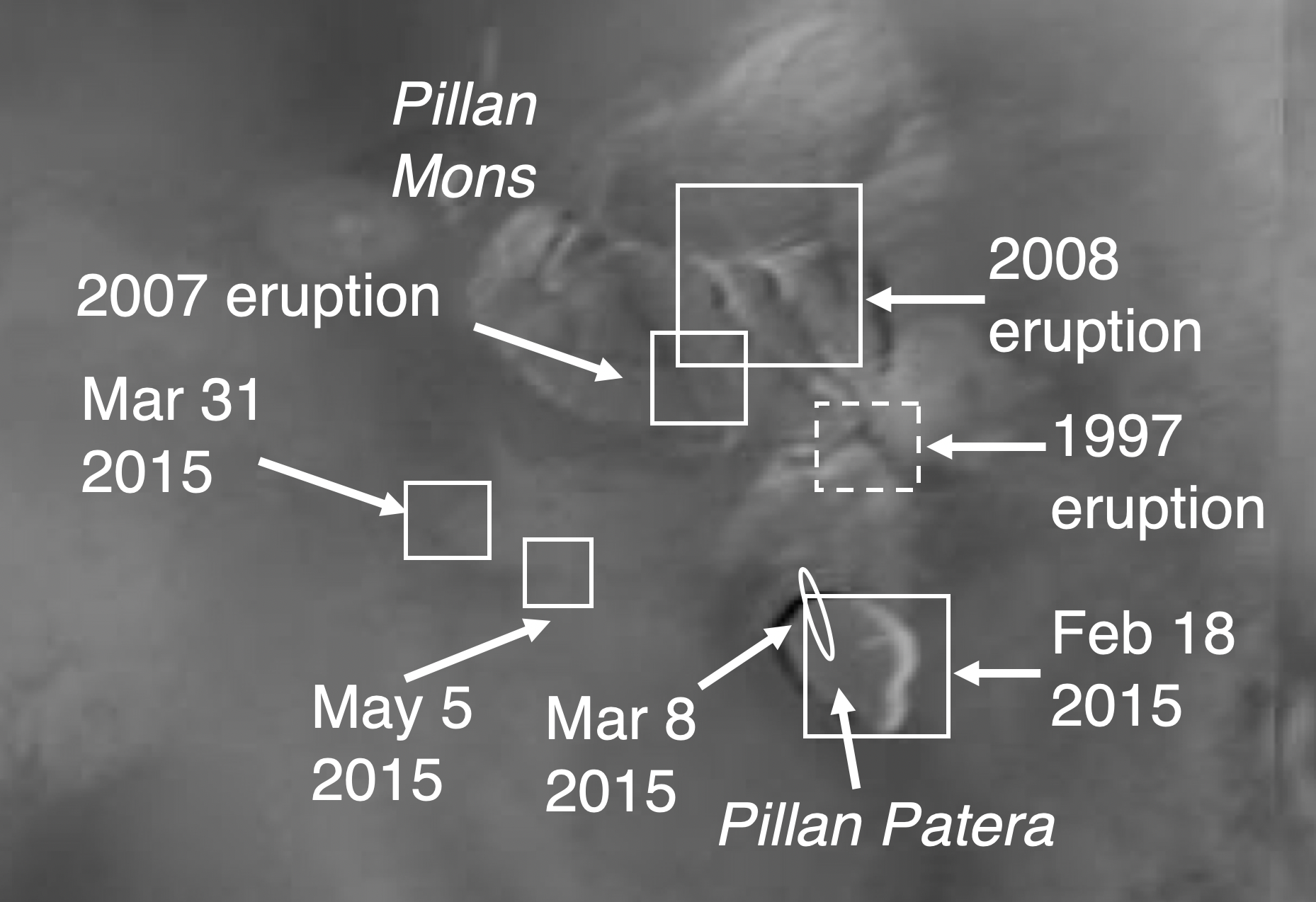}
\centering
\caption{Location of emission from Pillan Patera presented here on 2015 March 08, in the context of other measurements. The figure is modified from Fig 2 of \cite{depater2016}; the basemap is a combined \textit{Galileo-Voyager} mosaic with a spatial resolution of 1 km/pixel. \label{fig:pillanmap}}
\end{figure}

The occultation geometry for Pillan Patera was such that the ingress and egress directions were nearly the same, and the latitudinal resolution is therefore limited. Nevertheless, the best-fit emitting region as determined from the lightcurves (Figures \ref{fig:pillanlc} \& \ref{fig:pillanim}) falls precisely within the patera area. It is slightly to the west of the emitting location determined from the IRTF eruption detection, though is consistent with that location within uncertainties. The precise location and area of the best-fit model are given in Table \ref{tbl:results}. The location of emission as observed with AO imaging by \cite{depater2016} in the following months (March 31 and May 5) was 100--200 km to the west of the patera (see Figure \ref{fig:pillanmap}). The apparent westward progression of the peak emission from Feb to May suggests that emissions may be triggered along some kind of ``fault" line, or along subsurface conduit systems that are connected over hundreds of km. Note that hot spot locations derived from occultations and AO data rely on, respectively, ephemeris timing and limb-fitting, and the two methods may have small systematic differences. Earlier eruptions between 1997 and 2008 took place in the mountainous region 100--200 km north of the patera \citep[see Figure \ref{fig:pillanmap}; ][]{davies2001,lellouch2015}.

\subsection{Kurdalagon Patera}

Kurdalagon Patera was detected as an isolated hot spot by the \textit{Galileo} Near-infrared Mapping Spectrometer (NIMS) on several occasions between 1996 and 2000 \citep{lopes2004}. When detected, its 4.8 $\mu$m flux density was variable, ranging from 0.27 GW/$\mu$m/sr (1996 December 18, Galileo orbit E4) to 2.45 GW/$\mu$m/sr (1997 April 3, Galileo orbit G7), and 2.07 GW/$\mu$m/sr (2000 February 22, Galileo orbit I27; \cite{davies2012}).

Kurdalagon Patera was also detected as a hot spot forty times with Keck using adaptive optics from 2007 to 2018.  These observations yielded colour temperatures between 360 and 1290 K, and total power outputs between 47 GW and 1250 GW \citep{dkdp2016a,cantrall2018,dekleer2019}. Kurdalagon Patera was the likely site of two ``mini-outbursts'' – large, transient thermal spikes in the near-infrared that fall short of full ``outburst'' scale \citep{dkdp2016a} – in January 2015 and just three months later in April 2015.  Mini-outbursts may be lava fountain events, or the catastrophic overturn of the crust in a lava lake, based on their high thermal flux densities and rapid decay.  In either event, a relatively large incandescent area is revealed.  Single temperature fits to the Keck data yielded high temperatures of 1180 K and 1290 K, associated areas of 8.6 and 8.0 km$^2$, and total radiated fluxes of 950 GW and 1250 GW.  These two eruptions are noteworthy because they coincided with the brightening of the extended cloud of neutral sodium within the Jupiter system \citep{yoneda2015,dkdp2016a}, which is thought to be sourced from sodium-bearing molecules in Io's volcanic plumes. The 2015 event provides circumstantial evidence that some eruptions on Io’s surface are the source of variability in the neutral cloud, although not all activity seen in the infrared is associated plumes, and not all plumes appear to have NaCl and KCl \citep{depater2020,moullet2015}.

\begin{figure}[ht!]
\centering
\includegraphics[width=16cm]{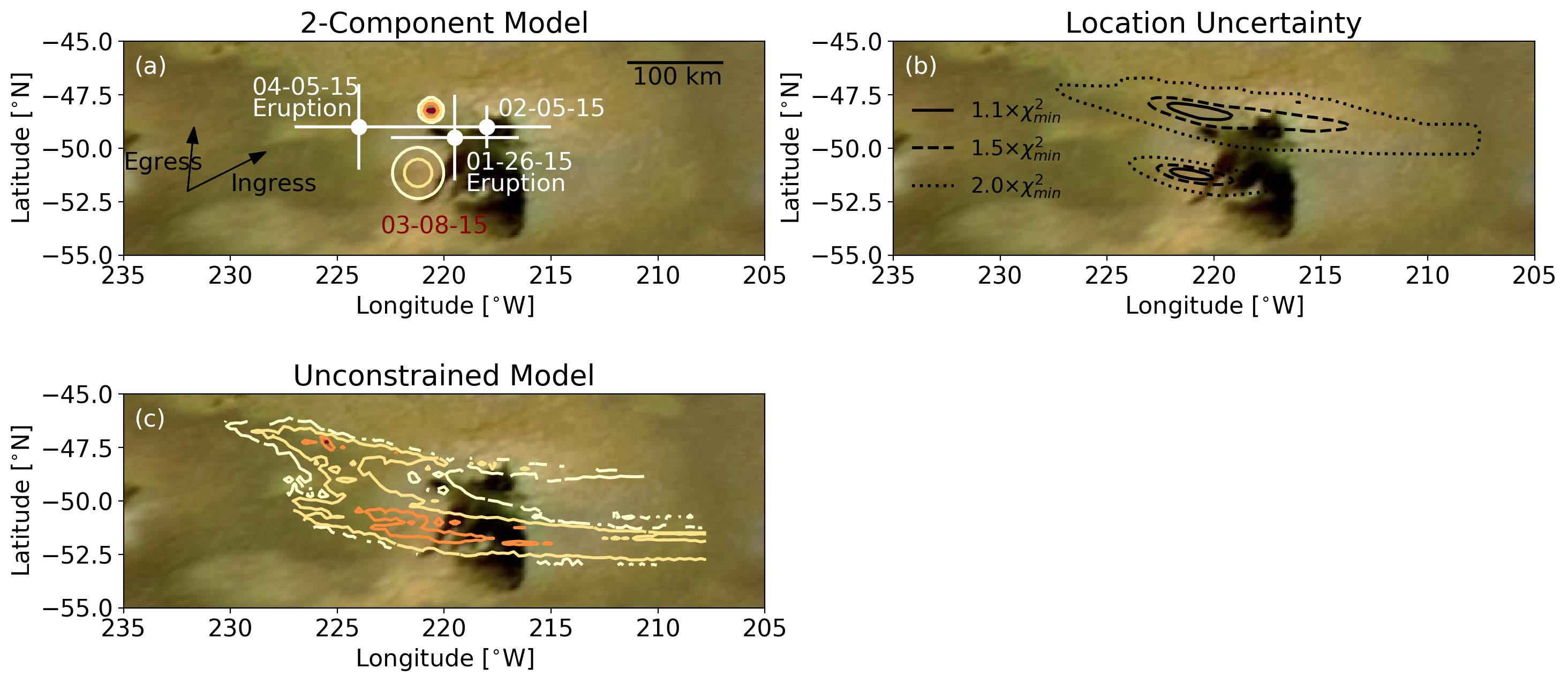}
\caption{Thermal emission from Kurdalagon Patera as derived from (a) a two-Gaussian model (our preferred model) and (c) from an unconstrained model, superposed 
 on a USGS \textit{Voyager/Galileo} mosaic. Contours are shown at 10, 25, 50, and 90\% of the peak emission in each model, together with the locations of two eruptions in early 2015. The approximate local directions of movement of Europa's limb during ingress and egress are indicated. Panel (b) shows the fractional change to the $\chi^2$ value of the 2-component model when the location of each component is moved individually with all other model parameters fixed at their best-fit values.\label{fig:v3im}}
\end{figure}

\begin{figure}[ht!]
\centering
\includegraphics[width=16cm]{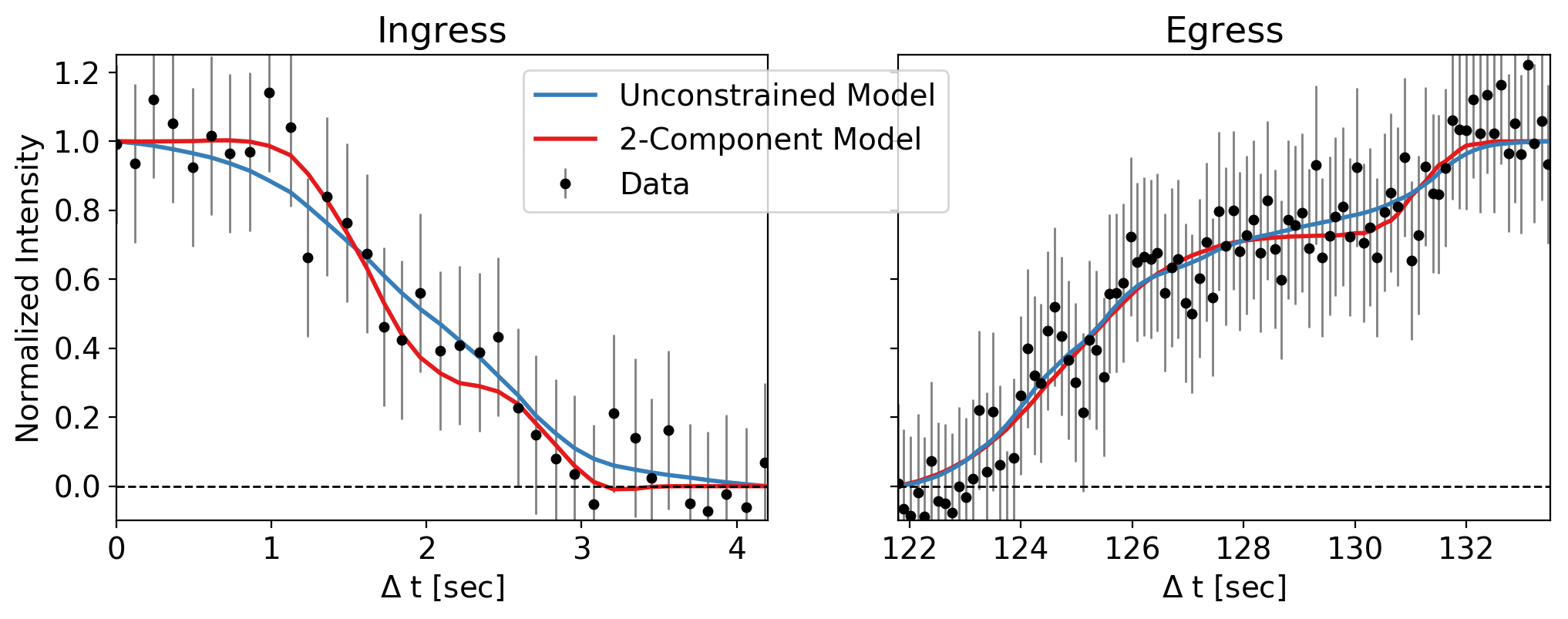}
\caption{Light curve of Kurdalagon Patera with best-fit models, corresponding to the emission distributions shown in Figure \ref{fig:v3im}. The evidence for two distinct emitting areas can be seen particularly in the egress light curve, with the two steps around $\Delta$t$=$122-126 seconds and 131-132 seconds. \label{fig:v3lc}}
\end{figure}

The LBTI data provide more precise locations for the thermal sources.  The location of the January 2015 eruption is consistent with the location of the patera itself, which was mapped by spacecraft. However, in several detections from March through April of 2015, including detection of a second large eruption on April 5, the emission appeared to be centered to the west of the patera around 223-225$^{\circ}$W. The emission detected by the LBT during the mutual event is centered on the west edge of the patera, consistent with both locations (Figure \ref{fig:v3im}). Models fit to the LBT light curve also indicate the presence of two emitting areas offset in latitude, as shown in Figure \ref{fig:v3im}. The evidence for this can be seen in the two steps in the egress light curve (Figure \ref{fig:v3lc}) occurring around $\Delta t$ of 124-126 and 131-132 seconds. Emitting areas this closely spaced would not be resolved with Keck, where the latitude uncertainty is typically a couple of degrees near Kurdalagon Patera (which is always foreshortened as viewed from Earth because of its latitude). The locations of the two components are indicated as stars on the image of Io in Figure \ref{fig:obsim}, which shows their proximity when projected to the viewing geometry. The locations, sizes, and flux densities of the two components are tabulated in Table \ref{tbl:results}.

\begin{figure}[ht!]
\centering
\includegraphics[width=20cm]{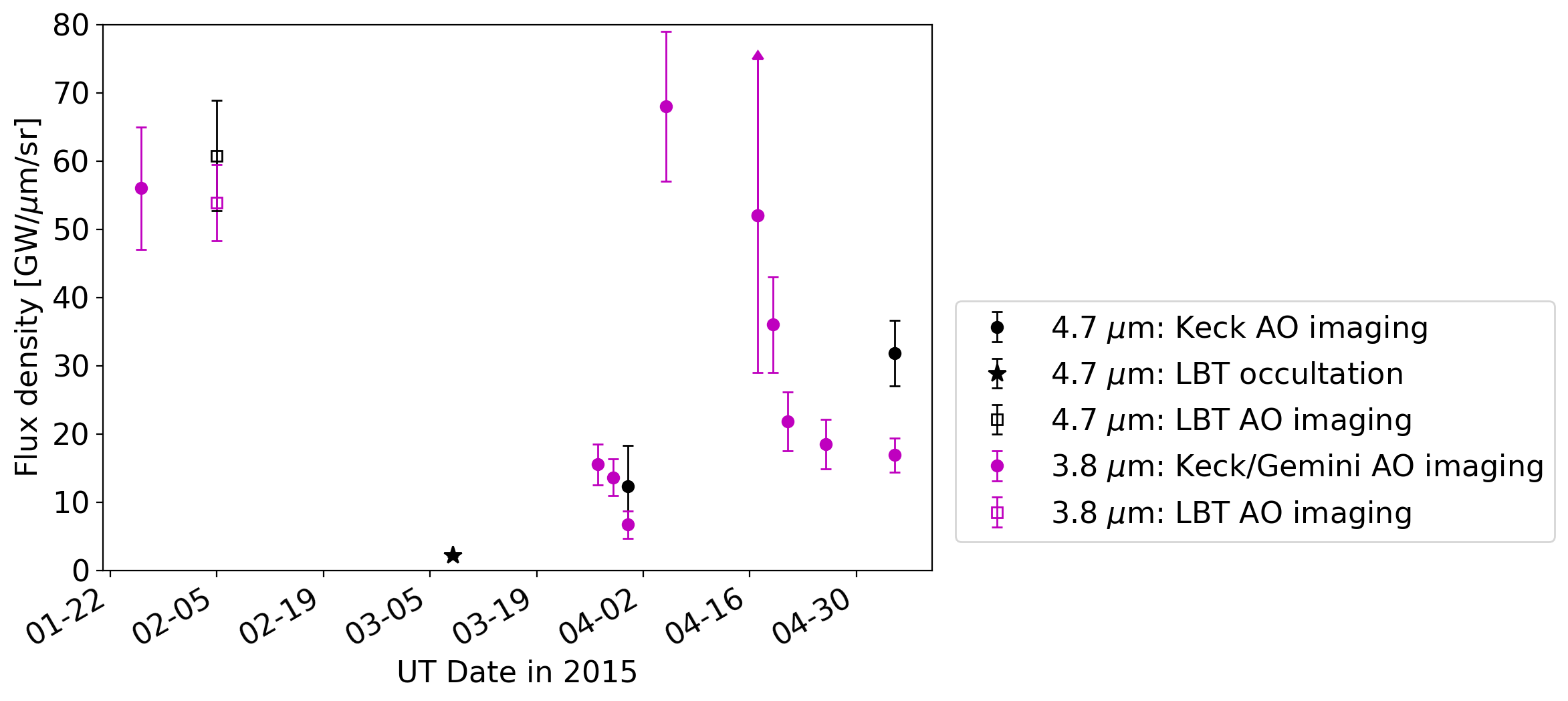}
\caption{Thermal emission from Kurdalagon Patera over time during its eruptions in early 2015. The LBT AO measurements are derived from the imaging dataset as presented in this paper, and all Keck and Gemini measurements are taken from \cite{dkdp2016a}. The uncertainties on the LBT occultation datapoint are smaller than the marker size. \label{fig:kurd_timeline}}
\end{figure}

The occultation observation finds a lower measured flux density than any other detection during that interval (see Figure \ref{fig:kurd_timeline}), indicating that these observations were made during a period of quiescence between eruptions. At the same time, the emitting areas derived from the LBT observations (45 and 16 km Gaussian FWHM) are much larger than the emitting area at eruption initiation (8.6 km$^2$; \cite{dkdp2016a}), suggesting that the LBT observations are sensing emission from a cooling lava flow following the January eruption. When the first observations of Kurdalagon Patera following these March 8th observations were made at the end of March, the volcano was brighter and decreasing in brightness, suggesting that a third eruption occurred but was missed between the January and April eruptions. 

To add context to the progression of the spring 2015 eruption at Kurdalagon Patera, we also analyzed adaptive optics images obtained with the LBT on 2015 February 05. The measured flux density of Kurdalagon Patera was 60.8$\pm$8.1 GW/$\mu$m/sr at M-band and 53.9$\pm$5.6 GW/$\mu$m/sr at L-band on that date, completely consistent with the brightness at the initial eruption detection on January 26th (Figure \ref{fig:kurd_timeline}). Note that the geometric foreshortening correction that was applied to these measurements (see Section \ref{sec:imaging}) would overestimate the emission at high emission angle if lava fountaining were contributing significantly to the emission. The best-fit location of 49$\pm$1$^{\circ}$S 218$\pm$3$^{\circ}$W derived from our Feb 5th data is also consistent with the emission location on January 26 (Figure \ref{fig:v3im}). These results together suggest that the eruption remained at a sustained high energy output for at least 10 days. The L and M band measurements together indicate a temperature and emitting area of 630$\substack{+130 \\ -90}$ K and 150$\substack{+230 \\ -100}$ km$^2$, cooler than the temperatures measured during the eruption's peak and consistent in area with the smaller occultation-detected hot spot.

\subsection{Ulgen Patera \& N Lerna Regio}

Ulgen Patera is a dark-floored patera with a floor area of 1598 km$^2$ \citep{williams2011,veeder2011}. It was identified in \textit{Voyager} data as a volcanic hot spot \citep{pearl1982,mcewen1985}, although it was combined with Babbar Patera and Svarog Patera into a single hot spot by \cite{mcewen1989} in an analysis of \textit{Voyager} data. \textit{Galileo} NIMS detected thermal emission of 500 GW from Ulgen Patera from an area of $\sim$1600 km$^2$ at $\sim$273 K \citep{veeder2012}, and it appeared as a stable thermal source in four NIMS observations in 1996 and 1999, with M-band flux densities between 4.2 and 6.3 GW/$\mu$m/sr.   

\cite{dkdp2016a} detected emission near Ulgen Patera and in N Lerna Regio on several occasions in early 2015. The emission from near Ulgen Patera was consistently observed to originate from an area to the northwest of Ulgen Patera, more consistent with the location of the nearby patera PV59. Located at 223$^{\circ}$W 38.3$^{\circ}$S, PV59 is also a dark floored patera \citep{radebaugh2005}.  In \textit{Voyager} imagery\footnote{https://astrogeology.usgs.gov/search/map/Io/Voyager-Galileo/Io\_GalileoSSI-Voyager\_Global\_Mosaic\_ClrMerge\_1km} PV59 was the likely source of a red surface pyroclastic deposit.  It is most probable that this deposit, like others seen on Io, is composed of short-chain sulfur allotropes \citep[e.g.,][]{mcewen1998}, indicating that active volcanism and the discharge of lava and associated volatiles was taking place during the \textit{Voyager 1} encounter in 1979.  

Emission from N Lerna Regio was detected from Keck and Gemini with a best-fit location of 57.3$\pm$1.2$^{\circ}$S 288.8$\pm$5.3$^{\circ}$W \citep{dkdp2016a}. Emission from both sites was detected on 2014 December 02 and on 2015 March 31, bracketing the LBT observations, and both volcanoes were very faint in March 2015 in Keck observations. 

\begin{figure}[ht!]
\includegraphics[width=14cm]{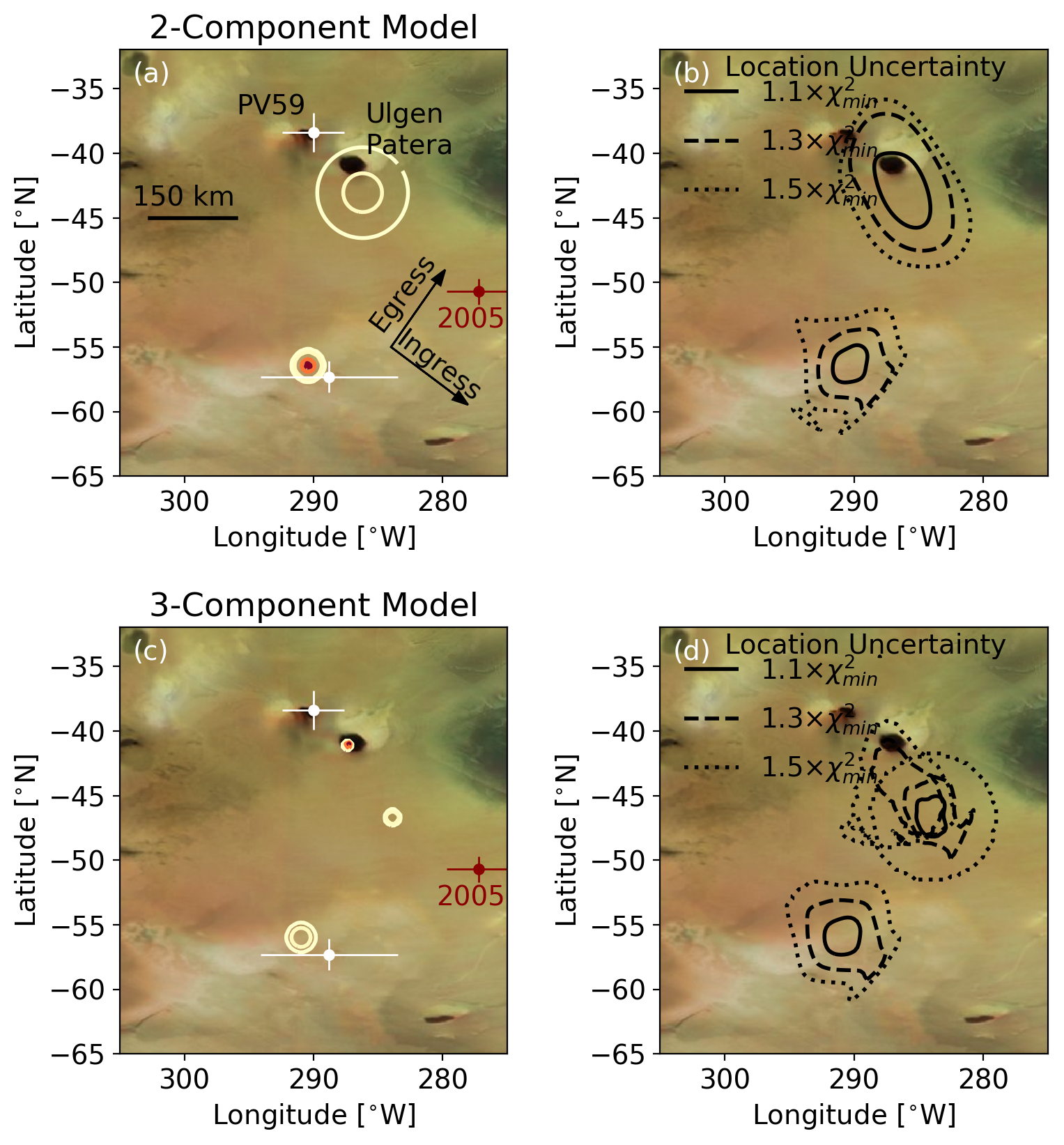}
\caption{Thermal map of Ulgen Patera/N Lerna Regio for  models that include (a) two, and (b) three Gaussian emitting areas, superposed on a USGS \textit{Voyager/Galileo} mosaic. The two paterae seen near -40$^{\circ}$N 290$^{\circ}$W are Ulgen Patera to the east and PV59 to the west. The approximate local directions of movement of Europa's limb during ingress and egress are indicated, and the white points represent the best-fit locations of the two hot spots observed in this vicinity in the \cite{dkdp2016a} adaptive optics dataset (both spots were observed on 2014 December 02 and 2015 March 31). The 2005 point is a detection from 2005 March 31 of a location labeled `S. of Babbar Patera' \citep{depater2016b}. The contours correspond to 1, 5, 50, and 90\% of the peak emission in each model. Panels (b) and (d) show the fractional change to the $\chi^2$ value of the 2-component and 3-component models respectively, when the location of each component is moved individually with all other model parameters fixed at their best-fit values.\label{fig:v4im}}
\end{figure}

\begin{figure}[ht!]
\includegraphics[width=16cm]{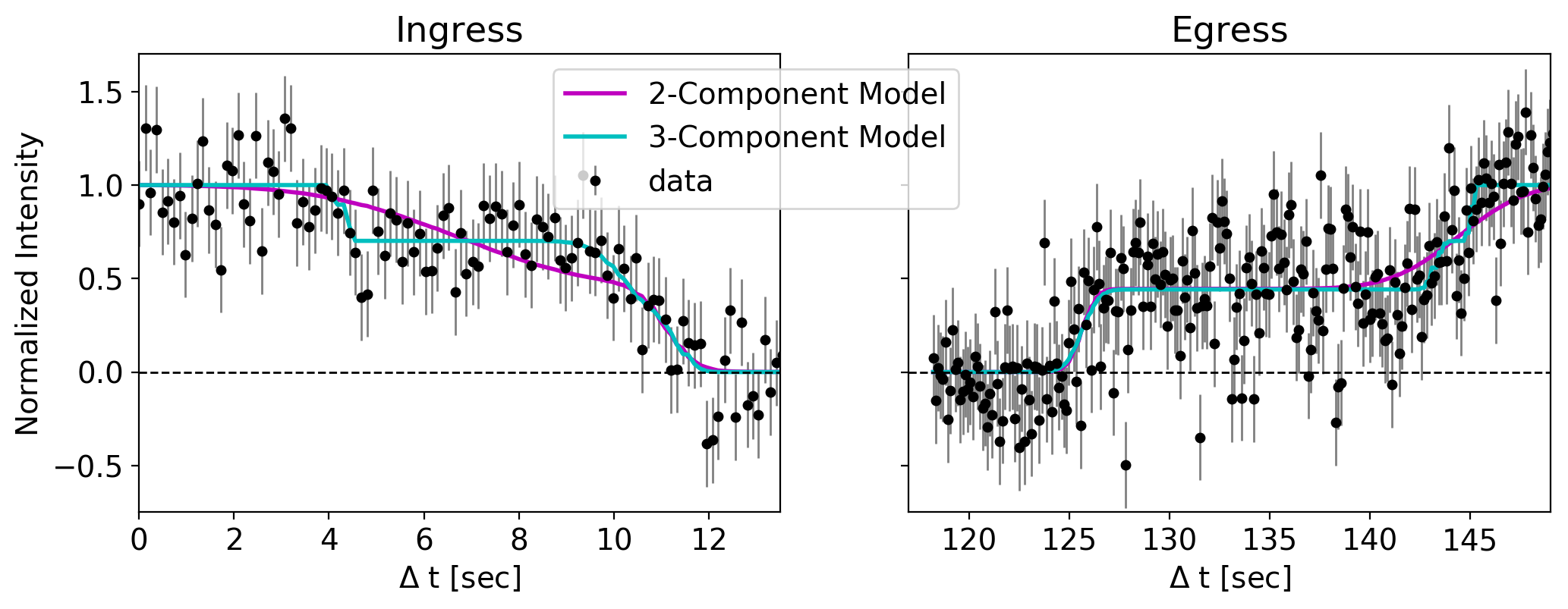}
\caption{Light curve of Ulgen Patera/N Lerna Regio with best-fit models, corresponding to the emission distributions shown in Figure \ref{fig:v4im}. \label{fig:v4lc}}
\end{figure}

Figures \ref{fig:v4im} and \ref{fig:v4lc} show the LBT light curve for this region and the best-fit intensity distribution for models that include two and three Gaussian emitting areas. Both models localize emission from N Lerna Regio to near 56$^{\circ}$S 290$^{\circ}$W, consistent with the location where emission was detected from Keck during the surrounding time period (also shown in Figure \ref{fig:v4im}). 
The northern emitting component takes the form of one broad emitting area or two compact emitting areas depending on how many components are allowed in the model. The 3-component model finds a better qualitative fit to the light curve, but both models provide reasonable fits given the uncertainties. In the 2-component model, the broad emission is centered a few degrees southeast of Ulgen Patera, whereas the 3-component model finds emission from directly within Ulgen Patera and from a second emitting component at 47$^{\circ}$S 284$^{\circ}$W. The locations of the three components are indicated as stars on the image of Io in Figure \ref{fig:obsim} and given alongside the sizes and flux densities of the emitting regions in Table \ref{tbl:results}. In either model, the emission is somewhat offset from the location detected by Keck and Gemini observations \citep{dkdp2016a}, which is centered on PV59 to the northwest of Ulgen Patera, indicating that both paterae remain active. The second component in the 3-component fit, southeast of Ulgen Patera, is midway between Ulgen Patera and the location of an eruption reported on 31 May 2005, which took place at 51$^{\circ}$S 277$^{\circ}$W and was labeled `S. of Babbar Patera' \citep{depater2016b}. At the time no prior eruptions had been seen at this location, and no associated surface features had been identified.

\section{Conclusions}\label{sec:conc}

We present an analysis of light curves of three distinct emitting hot spots on Io from LBTI observations taken on 2015 March 08 when Europa passed in front of Io. The emitting areas correspond to Pillan Patera, Kurdalagon Patera, and the vicinity of Ulgen Patera/PV59/N Lerna Regio. Pillan Patera was observed during the waning phase of an eruption, and the measured flux density matches the time evolution of the eruption derived from previous and subsequent measurements. The emitting region is centered to the northwest of the eruption site as determined three weeks earlier, and is oriented in the same direction as the location of the emission observed $\sim$three weeks later by other telescope facilities. Emission at Kurdalagon Patera originates from two sources, both consistent in location with emission detected $\sim$6 weeks prior to our observations, and with emission detected two months later. Previous measurements could not distinguish between a single or double source. Emission from the Ulgen Patera--N Lerna region likely consists of three small sources, close in position to emission detected on 2 December 2014 and 31 March 2015 from other facilities. One emission region is located at Ulgen Patera, another in N Lerna Regio in the vicinity of past detections, and a third in between these two regions.

The resolved structure of the thermal emission from Io's hot spots that is revealed here by the Large Binocular Telescope during a satellite occultation of Io adds a new dimension to the analysis of previous telescope imaging data, and most spacecraft data, where Io's hot spot sizes are sub-pixel. Occultation datasets have long been a powerful tool for localizing Io's eruptions; the combination of mutual occultation events with adaptive optics removes the degeneracy in determining emission location, and allows for more and fainter volcanoes to be measured and substructure within hot spots to be mapped. Adaptive optics mutual occultation datasets, as demonstrated previously at Loki Patera and now at other volcanoes, can therefore collect data previously only obtainable from spacecraft making close Io fly-bys. Though the opportunities to make such observations are rare, they provide an unprecedented level of detail on the thermal structure of Io's volcanic sources.

\clearpage

\acknowledgments

The LBT is an international collaboration among institutions in the United States, Italy and Germany. LBT Corporation partners are: The University of Arizona on behalf of the Arizona university system; Istituto Nazionale di Astrofisica, Italy; LBT Beteiligungsgesellschaft, Germany, representing the Max-Planck Society, the Astrophysical Institute Potsdam, and Heidelberg University; The Ohio State University, and The Research Corporation, on behalf of The University of Notre Dame, University of Minnesota and University of Virginia. The
LBT Interferometer is funded by NASA as part of its Exoplanet Exploration
program. The LMIRcam instrument is funded by the US NSF through grant NSF
AST-0705296. CEW acknowledges support from NASA PAST grant 80NSSC19K0868.

\bibliography{IoOccultation}{}
\bibliographystyle{aasjournal}

\end{document}